# Investigating the Investment Behaviors in Cryptocurrency

Dingli Xi
Master of Economics and Public Policy
School of Economics
University of Queensland
St. Lucia, QLD, 4072, Australia
d.xi@uq.net.au

Timothy Ian O'Brien
Bachelor of Politics, Philosophy and Economics (Honours)
School of Economics
University of Queensland
St. Lucia, QLD, 4072, Australia
timothy.obrien4@uq.net.au

Elnaz Irannezhad (Corresponding Author)
Postdoctoral Research Fellow
Australian Institute of Business and Economics
University of Queensland
St. Lucia, QLD, 4072, Australia
e.irannezhad@uq.edu.au

## Highlights

- The significant factors of the choice of investment in cryptocurrency include age, gender, education, occupation, and previous investment experience.

- Chinese and Australian investors rank the ICO attributes differently.

- The deterrence factors, and investment strategies vary between Chinese and Australians.

## Abstract

This study investigates the socio-demographic characteristics that individual cryptocurrency investors exhibit and the factors which go into their investment decisions in different Initial Coin Offerings (ICOs). A web-based revealed preference survey was conducted among Australian and Chinese blockchain and cryptocurrency followers, and a Multinomial Logit model was applied to inferentially analyze the characteristics of cryptocurrency investors and the determinants of the choice of investment in "cryptocurrency coins" versus other types of ICO tokens. The results show a difference between the determinant of these two choices among Australian and Chinese cryptocurrency folks. The significant factors of these two choices include age, gender, education, occupation, and investment experience, and they align well with the behavioural literature. Furthermore, alongside differences in how they rank the attributes of ICOs, there is further variance between how Chinese and Australian investors rank deterrence factors and investment strategies.

## 1. Introduction

With the emergence of the internet and its subsequent rapid growth in last few decades, the invention of the blockchain based on the modern internetwork seems to be an unsurprising product of this new era (Zhou et al., 2016). Since the creation of the first cryptocurrency, Bitcoin, in 2008 (Nakamoto, 2008), the cryptocurrency market has experienced exponential growth in the ten years following its inception. This in turn has turned cryptocurrency into one of the most attractive and fascinating buzzwords in the world. From less than 0.01 USD worth of bitcoin in 2010 to above 10,000 USD each in 2019, and from a single bitcoin to more than 1000 altcoins and crypto-tokens, cryptocurrency and blockchain have created more investment myths than anyone could have imagined.

The emergence of the underlying blockchain technology of decentralization and distributed ledgers has undoubtedly begun to bring revolutionary changes across many fields. A relatively recent and novel financial innovation is the so-called initial coin offering (ICO). The ICO, which relies on blockchain technology, is a niche form of crowd-funding used by blockchain startup ventures to launch a business based on distributed ledger or blockchain technology (Chanson, Gjoen, Risius, & Wortmann, 2018). Similar to Initial Public Offerings (IPOs), which the term ICO derives from, ICOs allow a company to issue transferable tokens to the general public which can be traded on open markets such as a cryptocurrency exchange. Unlike traditional fundraising mechanisms, the issued tokens can contain variable properties in representing their values, and this is most often the access to the firm's products and services (Sehra, Smith, & Gomes, 2017; Tapscott and Tapscott, 2017). The other difference is that participants in an ICO exchange cryptocurrencies that may or may not have rights attached. The issued token does not convey the ownership in the start-up company but can bring certain rights such as voting. Presumed interest in cryptocurrency, which comes from a fear of missing out on an opportunity, has led to a 15.2% growth in ICOs, with approximately 11.5 billion USD raised in 2018[1].

Although the ICO is still a very recent phenomenon in fundraising compared to traditional financing, it has been increasingly used by companies as an easy way to raise funds. The existing literature studying ICOs has focused on the determinants of the success of ICOs, however the body of literature is relatively scarce in investigating the characteristics of cryptocurrency investors.

This study aims to use choice modeling to examine the characteristics of investors and their choices regarding the types of ICOs they invest in. By understanding what characteristics investors exhibit and what factors go into their investment decisions, this study aims to develop a better understanding of how firms, particularly those running ICOs, can better attract investors and thereby further the expansion of the technology's applications despite current perceptions.

---

[1] ICO Market Quarterly Analysis 2018. https://icobench.com/reports/ICO_Market_Analysis_2018.pdf

Accordingly, we conducted a revealed preference (RP) experiment where blockchain folks were asked about their investment behaviors, their ICO preferences, and other attitudinal factors relating to risk and ICO attributes. We also conducted the same survey in China (translated into Mandarin) so as to compare the similarities and differences of investor characteristics as well as the perceived cryptocurrency landscapes across both countries. The attributes that might be relevant to investment in cryptocurrency have been borrowed from the literature on behavioral theories, similar studies, and brainstorming with representatives of cryptocurrency investors. We have applied the Multinomial Logit model to model the choice of investment in cryptocurrency as well as choice of investing in coins versus tokens.

This study indicates that the gender, age, income level, background education, and job of individuals are among the determinants of cryptocurrency investment. Notably, this study finds evidence which aligns well with theories of risk-aversion in behavioral economics and finance.

The rest of paper is organized as follows. Section 2 describes the research background and literature review with regards to the determinant factors in cryptocurrency investment. The experiment, methodology, and model specifications are described in Section 3. The results of the estimation of the models are presented in Section 4. Section 5 presents the descriptive statistics about ICO attributes, investment strategies, and friction factors of investment. Finally, concluding remarks are presented in Section 6.

## 2. Literature review and research background

In this section, we summarize the literature into two main categories: investors' characteristics, and ICO attributes.

## 2.1. Investors' characteristics

Different characteristics of investors in financial markets determine different preferences in financial instruments and different levels of financial risk-taking. Accordingly, the research on the factors influencing investor behavior is essential for blockchain-related start-ups to be successful. However, the literature in studying investors behaviors and characteristics in cryptocurrencies is relatively scant. There are only a few noteworthy examples of experimental studies and descriptive snapshots of cryptocurrency investor characteristics and ICO attributes.

Foley and Lardner LLP (2018) investigated investor perceptions of cryptocurrencies and found that regulation to reduce uncertainty is key for investors going forward. A survey from Encrybit[1] provides descriptive statistics from 161 countries on the impacts of several variables on cryptocurrency investment, namely token exchanges, risk management, news, and social networks.

Another survey[2] on Reddit presents a descriptive snapshot of investor characteristics from age and gender through to portfolio size and ideological justifications. A study by Mahomed (2017) is the only study using inferential statistical analysis where they investigated consumer adoption of cryptocurrencies through the lens of behavioral biases. Other than the descriptive snapshots shown in Exhibit 1, there is no study about the characteristics that may play a role in cryptocurrency investments.

---

[1] https://medium.com/@enbofficial/encrybit-cryptocurrency-exchange-evaluation-survey-2018-global-analysis-insights-b53c16abb106

[2] https://www.investinblockchain.com/reddit-cryptocurrency-survey/

Exhibit 1: Factors in the existing cryptocurrency surveys

| Survey | Age group | Gender | Education | Work Status | Income | Risk-Aversion | Behavioral Traits | Familiarity with ICO blockchain | Token type preference | Deterrents from investing | Important token traits | Risks of crypto | Future expectations of industry |
|---|---|---|---|---|---|---|---|---|---|---|---|---|---|
| Foley1 | | | | ✓ | | | | | | ✓ | | ✓ | |
| Reddit2 | ✓ | ✓ | ✓ | ✓ | | | | ✓ | | | | | |
| SharesPost3 | | | | | ✓ | | | | | | | ✓ | ✓ |
| Encrybit4 | ✓ | ✓ | | | | | | | | | | | |
| Circle5 | ✓ | ✓ | | | | ✓ | | | | | | | ✓ |
| Clovr6 | ✓ | ✓ | | ✓ | | | | ✓ | | ✓ | | ✓ | |
| Aicpa7 | ✓ | ✓ | | | | ✓ | | ✓ | | | | | |
| ING8 | ✓ | ✓ | | | | | | ✓ | | | | | ✓ |

When reviewing the literature, we also looked at the findings obtained from the traditional financial market in understanding investors characteristics. For example, financial risk tolerance is found to be a fundamental dimension to explain investor behavior and has been widely studied in the financial literature (Bucciol and Zarri, 2015; Grable, 2000). Financial risk tolerance simply indicates the maximum tolerance and willingness of someone to take risks in making financial decisions. Fisher and Yao (2017) noted that in an efficient market, investors who take larger risks would expect higher returns. As such, investors with a higher level of risk tolerance are expected to hold assets with a significantly higher level of risk in order to obtain a higher return in the long run. As mentioned in the previous section, ICOs appear to be moving towards becoming efficient markets and consequently we assume that the literature of risk tolerance characteristics based on traditional financial markets are, to some extent, applicable to ICO and cryptocurrency markets.

---

[1] https://www.foley.com/files/uploads/Foley-Cryptocurrency-Survey.pdf

[2] https://www.investinblockchain.com/reddit-cryptocurrency-survey/

[3] https://sharespost.com/downloads/SharesPost_Cryptocurrency_and_Blockchain_Survey.pdf?nc=1

[4] https://medium.com/@enbofficial/encrybit-cryptocurrency-exchange-evaluation-survey-2018-global-analysis-insights-b53c16abb106

[5] blog.circle.com/2018/09/12/new-study-millennial-women-underrepresented-in-crypto-investing-opportunity/

[6] https://www.clovr.com/page/emotional-currency

[7] www.aicpa.org/press/pressreleases/2018/americans-say-volatile-markets-are-easy-way-to-make-profit.html

[8] https://think.ing.com/uploads/reports/ING_International_Survey_Mobile_Banking_2018.pdf

A vast literature indicates that men, on average, are more likely to have a higher risk tolerance when making financial decisions compared to women (Byrnes, Miller, & Schafer, 1999; Charness and Gneezy, 2012; Jacobsen, Lee, Marquering, & Zhang, 2014; Lemaster and Strough, 2014; Neelakantan, 2010; Sung and Hanna, 1996). Deo and Sundar (2015) confirmed this hypothesis in the investment behaviors displayed in the Indian stock market and suggested that this might be due to the different gender behaviors. They found that men are more likely to have a higher risk tolerance level because they are relatively more active investors and make more investment decisions on a daily basis. Almenberg and Dreber (2015) argued that financial literacy could explain the significant gender gap in the stock market, suggesting that women are less likely to participate in the stock market due to a lower level of financial literacy. Fisher and Yao (2017) suggested that the gender gap of risk tolerance mainly comes from income uncertainties. They observed that women have less yearly income compared to men and consequently may need to keep a larger portion of money in accounts with low returns (low risk) to bare the possible negative income shocks. Barber and Odean (2001) links this difference to psychology and argues that men are more prone to overconfidence than women.

Furthermore, literature on the correlation between age difference and risk-taking activities is controversial. Some found that older individuals have a higher risk tolerance than younger individuals (Morin and Suarez, 1983). Others found that younger age groups were, on average, willing to take risks at a much higher frequency than older age groups which were at or near to retirement age (Finke and Huston, 2003; Grable, 2000). Wang and Hanna (1997) studied the U.S. households' risk tolerance and found that the proportion of risky assets held by households increased as people aged, indicating a higher risk tolerance amongst the elderly. They suggested that this may be due to the limited financial resources or income of youths, causing younger people to have less capacity for enduring short-term investment losses.

People with higher incomes are willing to take more financial risks than those with lower incomes (Carducci and Wong, 1998; Finke and Huston, 2003; Grable, 2000; Roszkowski and Grable, 2010). Similarly, Morin and Suarez (1983) studied the risky assets held by Canadian individuals and concluded that risk tolerance, to some extent, depends on the levels of net worth. They found that risk tolerance decreases as age increases for households which have low levels of net worth. In contrast, for the households which have higher levels of net worth, the risk tolerance increases with age. However, Riley and Chow (1992) observed a different pattern of holding assets in American households. They found that household risk tolerance is positively correlated with age until they reach 65 after which there is a negative correlation. Their results, however, confirmed the discussions above that risk tolerance is related to net wealth. They showed that risk aversion decreased significantly as an individual's wealth rose into the top 10% of the population.

We can, therefore, reasonably concluded that women are more risk-averse than men, the young are more risk-seeking than the old, wealthier individuals manifest a greater willingness to invest in riskier assets, and the poor are risk-averse (Ndirangu, Ouma, & Munyaka, 2014).

### *2.2. ICO attributes*

Coins and tokens are two terms used to describe the unit of cryptocurrency value and are used to transfer value as a means of payment similar to money. However, tokens have a wider functionality than coins such as delivering value to investors beyond speculative returns and holding votes on key business decisions such as technical or protocol changes to the platform.

Cryptocurrencies can be categorized into: (i) 'currency' coins which grant the right to another digital currency to holders; (ii) 'asset/platform' tokens which grants the right to a promised future cash flow linked to an underlying business; (iii) 'utility' tokens that grant the right to access a product or service that the startup provides usually at some future point; and (iv) 'security' tokens which confer rights to an equity stake in a business and are used to invest or trade on the market. Based on the type of use-cases, business, and platform, these categories can be further broken down, with, for example, the website 'icobench.com' displaying 28 categories, with platform, cryptocurrency, and business services currently amongst the top three by quantity (Exhibit 2).

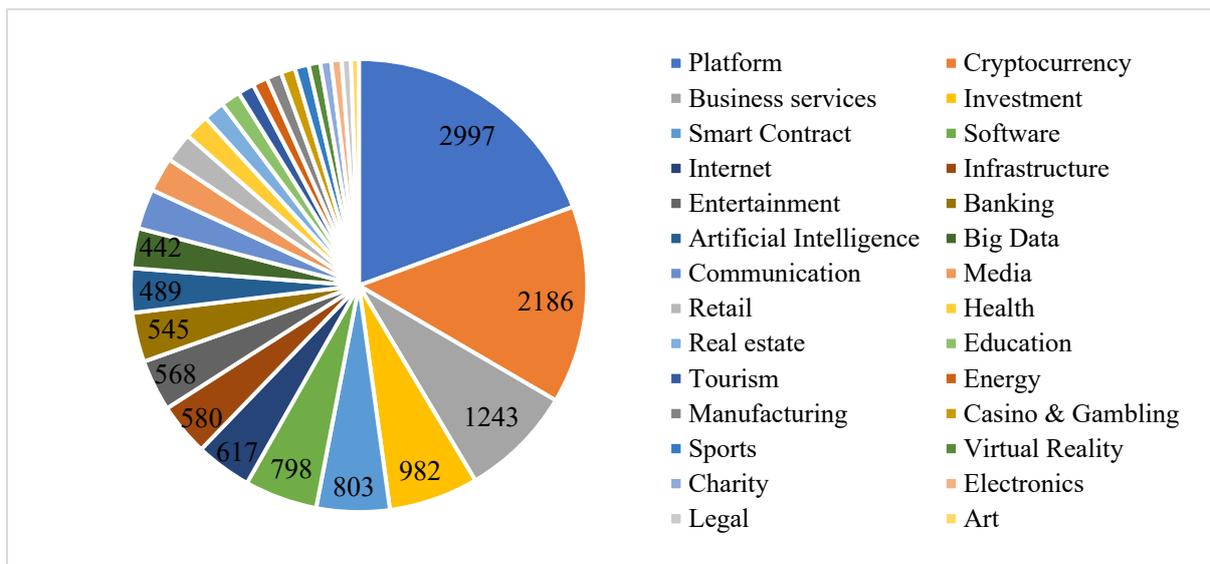

Exhibit 2: The categories of ICOs (source: icobench.com, retrieved in May 2019)

Overall, these different types of cryptocurrencies can be grouped into two major categories of coins and tokens. Coins present a group of alternative coins as an alternative to fiat money while tokens present the cryptocurrencies associated with their ~~the ther~~ use-cases of blockchain rather than the concept of cryptocurrency.

The main body of scientific literature focuses on what makes an ICO successful, with results ranging from the quality of its whitepaper through to how well it has built upon network effects via communities and social media. Information asymmetry is always the core problem influencing investor trust as well as the success of any form of fundraising activities, and this has been widely studied in the literature (Chod and Lyandres, 2018). This is a concept describing the fact that fundraising companies have greater material knowledge than potential investors in the actual quality of their company or project (Firoozi, Jalilvand, & Lien, 2017). In the traditional fundraising activities such as IPOs and venture capital, this information asymmetry can be effectively alleviated by comprehensive supervision and regulation, as disclosure regulation and enforcement are considered the foundations of well-functioning capital markets (Bourveau, George, Ellahie, & Macciocchi, 2018). However, in the context of cryptocurrency, ICOs are not well regulated in most jurisdictions around the world, and this results in a massive amount of misconduct and scams (Barsan and LL.M., 2017). The ICO ventures know more about the quality of their project than those potential investors, putting ICO investors in an informational disadvantage. The ICO information asymmetry has been widely observed in the literature (Felix, 2018; Momtaz, 2018).

Underpricing is a well-observed phenomenon in the process of IPOs where the initial price of the stock is set to be much lower than its intrinsic value. Information asymmetry is the most significant and accepted explanation of underpricing. Felix (2018) found robust evidence that underpricing is even more common among ICOs compared to IPOs, indicating a significant level of information asymmetry. Due to information asymmetry and the information gap, investors are easily deceived. Howell, Niessner, & Yermack (2018) suggested that the liquidity and trading volume for tokens would be higher if the issuers took steps to reduce the information asymmetry.

However, if a high level of information asymmetry continues to exist in the long-run, it could turn the ICO market into a market for lemons (Sehra, et al., 2017). "The market for lemons" theory is first explored by economist Akerlof (1970), which explains how the quality of goods traded in a market will leave only bad quality product by studying the used car sales market. Sehra, et al. (2017) concluded that the ICO market is suffering from the "lemon" condition. By studying the five characteristics in defining a lemon market, they proposed that the ICO market contains a high level of information asymmetry, skewed incentives, lack of disclosure framework, low-quality product, and lack of assurance, all of which are common in a "lemon" market.

Due to the absence of regulation or due diligence, investors can only rely on the information that the company wants them to know. "Lemon" theory suggests that investors in the market, therefore, are only willing to pay a portion of the value they think is appropriate for ICOs, leading to a vicious circle resulting in only the low quality and fraudulent ICOs left in the market. As low-quality ICOs increase in number on the market, the lemon market will eventually squeeze out all current and potential investors. Therefore, in order to gain more investors and build investor trust in the market, it is necessary to reduce the information asymmetry.

A higher level of information asymmetry generally leads to a greater need for signaling (Allen and Faulhaber, 1989). Signaling theory, first proposed by Spence (1978), can be applied to reduce the information asymmetry between two parties, assuming that the issuer knows more than the investor and is willing to signal the actual value. Signaling theory suggests a very efficient way of describing issuer behavior, fundraising activities, and reducing the information asymmetry (Connelly, Certo, Ireland, & Reutzel, 2010). Generally, because of the existence of information asymmetry between ventures and investors, the ventures need to disclose credible information in order to verify the credibility and validity of their project (Kromidha and Robson, 2016). Under a high level of information asymmetry, the behaviour of potential investors will depend on how the issuers send signals in order to encourage more investors (Courtney, Dutta, & Li, 2017).

A technical white paper could be a sufficient signal in ICOs (Fisch, 2019). The white paper is an essential component of a venture's ICO campaign which can be traced back to Bitcoin (Nakamoto, 2008). It is generally a document used by ventures to provide information which they deem necessary to show publicly (Barsan and LL.M., 2017). A white paper usually discloses and outlines the token distribution, business model, and other additional information related to the project. It is widely recognized as the main source of information for potential investors. A white paper should, therefore, be a reliable and informative tool for potential investors.

On the other hand, Chod and Lyandres (2018) argued that unlike venture capital or other fundraising activities, the white paper is not a reliable tool for investors due to lack of regulations and investment protection schemes. As discussed by Felix (2018) cryptocurrencies are a new unknown and unregulated form of investing in which only a few people know the true characteristics of a project. Fisch (2019) further suggests that ICOs typically occur in the early stages of a venture's life cycle and the tokens often do not have any counter value or real-world usage at the time of the ICO. Additionally, Zetzsche, Buckley, Arner, & Föhr (2017) suggest that more than half of the white papers on the market do not disclose project contact details, and an even higher number of white papers do not provide any underlying suitable legal support, money, or funds supervision and compelling audit. Momtaz (2018) also suggested that white papers are often designed on a euphemistic way for luring investors, and consequently investors cannot trust the white paper due to a lack of proven track record of ICO start-ups and lack of corporate governance and regulations.

Instead of a white paper, Adhami, Giudici, & Martinazzi (2018) argued that the success of an ICO mostly depends on the source code shared on the internet. They have found in studying various ICOs that the probability of the success of ICOs is unaffected by the availability of white papers, but is significantly positively affected by the availability of the source codes. They explained that white papers revealed by ventures have different information qualities, and the mere presence of one such document as an attachment to the ICO announcement is not particularly valued by potential investors, particularly due to lack of regulations and audits. By contrast, the informative power of source codes is very significant, providing a valuable and tangible measurement to analyze the intrinsic value of the ICO project.

Similarly, Fisch (2019) found that high-quality source code is associated with the success of an ICO, and suggested the source code is an objective feature for most investors in distinguishing different ICOs when making investment decisions. While one may argue that the quality of code is too technical to be understood by the majority of investors, the author argues that through the availability of source codes, investors are assisted by the technical reviewers, website news, and social media about the underlying technological capabilities of the project. Further, Bourveau, et al. (2018) reviewed 776 attempted ICOs and found that successful ICOs are more likely to reveal their source codes through GitHub disclosures or other publicly available code repositories. By contrast, ICOs which did not disclose their source codes on those platforms were likely to be the one which failed to meet fundraising goals. To explain this phenomenon, they argued that compared to "soft" or unaudited information disclosed in unverifiable ways, investors prefer more credible sources of information such as source codes.

A reliable scoring system could be another effective signal in ICOs, as suggested byBian et al. (2018) who found that a reliable credit rating system is necessary and urgent for ICOs. An appropriate and well developed ICO rating could prevent individual investors from losses by providing legality, transparency, and technical understanding to the market. Feng, Li, Lu, Wong, & Zhang (2018) also found a positive association between the ICO rating and the amount of funds raised. Boreiko and Vidusso (2019)again found that the probability of success in ICO fundraising campaigns is associated with extensive coverage in the ICO aggregators' lists. However, they noted that the rating data and results vary considerably across different rating websites, and suggested the investors should further investigate these rankings. Additionally, the majority of ICO rating mechanisms depend on the information revealed in white papers. Since many researchers argue that the white papers are not credible sources of information, the current ICO rating website or mechanisms which largely depends on the white paper might be biased, and rational investors might not take ICO ratings on the market as a trustworthy source of measurement.

According to this literature review, the potential relevant attributes of the investor's characteristics as well as ICO attributes were extracted to be investigated in the experiment. Accordingly, the behavioural literature highlights some investor characteristics including age, gender, education, occupation, and investment experience. The literature on ICO attributes also suggests a number of important factors, namely insider information, advertising, fundraising strategies, source code, white papers, and regulation.

## 3. Methodology

### *3.1. Experiment*

A revealed preference (RP) survey was developed to collect data on preferences towards the investment, and the type of investment. The survey was simultaneously distributed among Chinese and Australian cryptocurrency folks. The simultaneity of data collection ensured the robust comparison between these two groups without impacts of price fluctuation. Both surveys were distributed within cryptocurrency and blockchain related social network channels (e.g. Telegram, Slack, Wechat and Facebook) and a number of blockchain-based or cryptocurrency-enthusiast groups, meetups, and events.

Data was collected with a Qualtrics-created web-based questionnaire. The web-based survey ensured confidentiality through anonymous participation as well as randomizing the items to avoid survey fatigue effects. The use of a web-based survey creates a sample that may not necessarily be representative of the whole population and may over-represent younger and educated participants. However, the topic of cryptocurrency itself may also attract these audiences more than other groups since buy and sell actions are also made through social media or online.

The questionnaire included 29 questions and had an expected completion time of 10 minutes. The questions focused on the theorized hypotheses according to the prior research presented in the previous two sections. The questionnaire was divided into two sections. The first section involved personal data including gender, age, income level, job, and education. The second section investigated the ICO attributes, general attitudes, and perceptions about cryptocurrency and ICOs.

The investor's characteristics, ICO features, and other relevant attributes were derived from a review of the literature on behavioral financial investment, as outlined in the Section 2. The categorization of variables were mainly chosen based on the Census Statistics, as reported by the Australian Bureau of Statistics[1]. In order to compare the results of the Australian and Chinese datasets, we considered the same categories for the Chinese survey save for the income levels which were chosen according to the income quantiles in China.

---

[1] https://www.abs.gov.au/websitedbs/D3310114.nsf/Home/census

The administration of the Australian survey resulted in 130 responses in 3 weeks of which 107 completely answered all the questions. The administration of the Chinese survey resulted in 485 responses of which 295 were complete and could be used for model estimation. The summary of the characteristics of the participants in the Australian and Chinese surveys has been presented in the Appendix 1.

*3.2. Method*

Discrete choice models describe decision-makers' choices among alternatives. Choice models are estimated under the assumption that decision-makers are utility maximizers. The utility of alternative $i$ for decision-maker $n$ is expressed as the sum of a deterministic part $V_{ni}$ and a stochastic part $\varepsilon_{ni}$ (Train, 2009):

$$U_{ni} = V_{ni} + \varepsilon_{ni} = \boldsymbol{\beta_1} \boldsymbol{X_n} + \boldsymbol{\beta_2} \boldsymbol{Z_{ni}} + k_i + \varepsilon_{ni} \tag{1}$$

where $\boldsymbol{X_n}$ is a vector of characteristics of decision-maker $n$, $\boldsymbol{Z_{ni}}$ is a vector of attributes of alternative $i$ for decision maker $n$, vectors $\boldsymbol{\beta_1}$ and $\boldsymbol{\beta_2}$ of fixed parameters associated to $\boldsymbol{X_n}$ and $\boldsymbol{Z_{ni}}$, and $k_i$ is the alternative-specific constant that captures the average impact on utility of all factors that are not included in the model.

The behavioral model is therefore choosing alternative $i$ if and only if $U_{ni} > U_{nj}$, $\forall\, j \neq i$. Expressing the probability of choosing alternative $i$ as a Logit model results in the following probability (Train, 2009):

$$P_{ni} = Prob(U_{ni} > U_{nj}\ \forall j \neq i) = Prob(\varepsilon_{nj} - \varepsilon_{ni} \prec V_{ni} - V_{nj}\ \forall j \neq i) \tag{2}$$

Using the density $f(\varepsilon_n)$, this cumulative probability can be rewritten as (Train, 2009):

$$P_{ni} = \int_{\varepsilon} \sum_n \ln(I(\varepsilon_{nj} - \varepsilon_{ni} \prec V_{ni} - V_{nj}\ \forall j \neq i)) f(\varepsilon_n) d\varepsilon_n \tag{3}$$

Where *I* is an indicator function equaling *"one"* when the term in parentheses is true and *"zero"* otherwise. Assuming the unobserved portion of utility $\varepsilon_{ni}$ is distributed IID extreme value and a type of generalized extreme value, the eq. (3) has a closed form expression for this integral.

Considering we have not collected the data about the type of cryptocurrency, we do not have **$Z_{ni}$**. It should be noted that the characteristics of decision-makers do not vary over the alternatives, and since they can only enter the model if create differences in utility over different alternatives. Hence, one of these parameters in one alternative is normalized to zero and the estimated *$β_1$* is interpreted as the differential effect of that parameter on the utility compared to the base alternative. Lastly, the elements of the vectors *$β_1$* are estimated by the maximum likelihood function, using PandasBiogeme (Bierlaire, 2018).

### *3.3. Model specification*

This section presents the specifications of the choice of investment and the type of investment. From all the information collected, all potential explanatory variables were added progressively to test whether they were statistically significant at least at the 10% level and, if they were significant, they were retained in the model, otherwise they were removed. Finally, the best model specifications were determined after testing for all combinations of variables and checking for partial correlations. Notably, the model specifications presented in the following sub-sections only include the variables which were statistically significant at the 10% of confidence level.

#### *3.3.1. Model specification for the choice of investment in cryptocurrency*

After specification testing, the utility equations were specified as shown in eqs.*(4-6),* and eqs.*(7-9)* for the Australian and Chinese surveys respectively. It should be noted that the error terms are Gumbel distributed, and the equations present the deterministic parts of the utility functions.

$$V_{1n} = \beta_{1,constant} + \beta_{IncomeCat3}IncomeCat3_n + \beta_{JobBanking}JobBanking_n +$$
$$\beta_{Female}Female_n \tag{4}$$

$$V_{2n} = 0 \tag{5}$$

$$V_{3n} = \beta_{JobOther}JobEduEntrepreneur_n + \beta_{JobIT}JobIT_n +$$
$$\beta_{InvestEquities}InvestEquities_n + \beta_{MajorEBF}MajorEBF \tag{6}$$

Where $V_{1n}$, $V_{2n}$, and $V_{3n}$ are the utility of investing now, never investing, and intention to invest in the future among the Australian dataset. These models were specified for each decision-maker $n$ by considering $Female_n$ as indicating being female, $IncomeCat3_n$ as the variable encapsulating individuals who had an income of 1000 to 2500 AUD on average per week, $InvestEquities_n$ as the binary variable indicating individuals who had previously invested in equities, $JobBanking_n$ as the variable encapsulating individuals who were employed in the banking sector, $JobIT_n$ as the variable encapsulating individuals who were employed in the Information Technology sector, $JobEduEntrepreneur_n$ as the variable indicating individuals who are business owners, entrepreneurs or employed in education, $MajorEBF_n$ as the variable encapsulating individuals who have studied economics, business, or finance.

The best specifications for Chinese dataset are presented in eqs(7-9), where $V'_{1n}$, $V'_{2n}$, and $V'_{3n}$ are the utility of investing now, never investing, and intention to invest in the future among the Chinese dataset.

$$V'_{1n} = \beta'_{1constant} + \beta'_{InvestEquities}InvestEquities_n \tag{7}$$

$$V'_{2n} = \beta'_{Job\ Wholesale}Job\ Wholesale_n + \beta'_{Postgrad}Postgrad_n + \beta'_{Age18\_30}Age18\_30_n \tag{8}$$

$$V'_{3n} = \beta'_{Age40above}Age\ 40\ above_n \tag{9}$$

Here, the variable $Age18\_30_n$ indicates individuals within the age group of 18 to 30 years old, $Age40above_n$ as the variable encapsulating individuals within the age group of 40 and above, and *JobWholesale*ₙ as the variable encapsulating individuals who were employed in the wholesale sector, *Postgrad*ₙ as the variable indicating individuals whose highest level of education was postgraduate.

*3.3.2. Model specification for the choice of cryptocurrency*

As mentioned before, we have considered two type of cryptocurrenct investments namely coin and token investment. In the Australian survey, coins are among preferred options where 66.6% of the observations have invested in coins only or in addition to other types of tokens. In the Chinese survey, this percentage is 24.82%. After specification testing, the utility equations were specified as shown in eqs.*(10-11)*, and eqs.*(12-13)* for the Australian and Chinese surveys respectively.

$$V_{coin,n} = \beta_{constant} + \beta_{Age18\_30}Age18\_30_n + \beta_{IncomeCat4}IncomeCat4_n +$$

$$+\beta_{Freelancer}Freelancer_n + \beta_{JobWholesale}JobWholesale_n + \beta_{majorIT}MajorIT_n +$$

$$\beta_{CasualUnemployed}CasualUnemployed_n \tag{10}$$

$$V_{token,n} = 0 \tag{11}$$

Where *V*<sub>coin,n</sub> is the utility of investing in coins for respondent *n*, and *V*<sub>token,n</sub> the utility of investing in other ICO tokens, in the Australian survey. The parameter *Age18_30ₙ* indicates being the age between 18 and 30, *IncomeCat4ₙ* as the variable encapsulating individuals who had an income higher than 2500 AUD on average per week, *Freelancerₙ* as the binary variable indicating individuals who work as freelancers, *JobWholesaleₙ* as the variable encapsulating individuals who work in the wholesaling sector, *MajorIT*ₙ as the variable encapsulating individuals who studied information technology, computer science or software engineering, and *CasualUnemployedₙ* as the variable encapsulating individuals who are either unemployed or work on a casual basis.

The best specifications for the Chinese dataset are presented in eqs. (12-13), where $V'_{coin,n}$ is the utility of investing in a coin and $V'_{token,n}$ is the utility of investing in other ICO tokens.

$$V'_{coin,n} = \beta'_{constant} + \beta'_{InvestEquityFixed} InvestEquityFixed_n + \beta'_{Male} Male_n + \beta'_{Freelancer} Freelancer_n \qquad (12)$$

$$V'_{token,n} = 0 \qquad (13)$$

The variable $InvestEquityFixed_n$ indicates individuals who have invested in equities and/or fixed incomes, $Male_n$ as the variable indicating being male, and $Freelancer_n$ as the variable encapsulating individuals who work as freelancers.

## 4. Results

This section presents the results of the model for two choices of investment and the type of investment.

*4.1. Model Results of the Choice of Investment in Cryptocurrency*

Respondents were asked to reveal their investment behavior, and accordingly three alternatives have been modelled, namely the choice of investment in cryptocurrency up to now, the choice of never investing, and the intention of investing in future. In the Australian survey, 78.2% of the observations have invested in cryptocurrencies, 11.9% plan on investing in the future, and 9.9% have not invested and do not intend to do so. In the Chinese survey, these percentages are 29%, 27%, and 44% respectively.

Exhibit 3 presents the parameters found within both models, all of which were found to be significant at the 95% level of confidence. The Australian model provides evidence of interest in cryptocurrency investment by those within information technology jobs which may be expected given their experience in the field. Interestingly, those who have experience investing in equities were less likely to be interested in investing in the future. Further, females were more likely to be considering investing in the future or never investing, which may indicate they are more risk-averse as was expected from the literature. Those who were earning between 1000 AUD and 2500 AUD per week on average and those who have a job in banking sector were more likely to invest now.

In the Chinese survey, individuals who belong to the age group 18 to 30 years old would be expected to have more risk tolerance, and they are relatively more willing to invest in cryptocurrency. Additionally, individuals within the age group above 40 years old were more likely to invest in this market in the future. However, when comparing the goodness of fit, the significant parameters in the Chinese model cannot strongly explain the choice of investment while the Australian model yields relatively satisfactory results.

Exhibit 3: statistics of variables – Choice of Investment

| Alternatives | Australian survey | | | Chinese Survey | | |
| --- | --- | --- | --- | --- | --- | --- |
| | Parameters | Statistics Average | t - test | Parameters | Statistics Average | t - test |
| Alternative 1: Already invested or invest now | $\beta_{1,constant}$ | 2.04 | 4.56 | $\beta'_{1,constant}$ | | |
| | $\beta_{IncomeCat3}$ | 0.502 | 2.52 | $\beta'_{InvestEquities}$ | 1.01 | 3.07 |
| | $\beta_{JobBanking}$ | 3.16 | 16.2 | | | |
| | $\beta_{Female}$ | -0.695 | -2.27 | | | |
| Alternative 2: Never invest | | | | $\beta'_{JobWholesale}$ | 0.23 | 2.58 |
| | | | | $\beta'_{Postgrad}$ | -0.20 | -2.04 |
| | | | | $\beta'_{Age18\_30}$ | -0.23 | -1.81 |
| Alternative 3: Invest in future | $\beta_{JobEduEntrepreneur}$ | -4.18 | -52.6 | $\beta'_{Age40above}$ | 0.23 | 2.38 |
| | $\beta_{JobIT}$ | 2.63 | 3.76 | | | |
| | $\beta_{InvestEquities}$ | -1.68 | -2.29 | | | |
| | $\beta_{MajorEBF}$ | 0.35 | 1.65 | | | |
| **Number of parameters** | 8 | | | 6 | | |
| **Number of observations** | 102 | | | 292 | | |
| **Null LL** | -112.06 | | | -320.795 | | |
| **LL** | -51.10 | | | -304.565 | | |
| **Adjusted $\rho^2$** | 0.54 | | | 0.051 | | |
| **Akaike Information Criterion** | 118.22 | | | 621.129 | | |
| **Bayesian Information Criterion** | 139.22 | | | 643.190 | | |

*4.2. Model results for the choice of cryptocurrency*

In our survey, respondents were asked to reveal their preferred ICOs to invest in. Considering cryptocurrency investors often distribute their investments across different types of ICOs, respondents could choose multiple options across seven types of ICO tokens and coins, namely: (1) currency coins; (2) trading tokens such as trading energy, water, and healthcare; (3) commodity tokens such as supply chain of asset or goods (i.e. diamond, food, and property); (4) identity management tokens; (5) social platform tokens such as Telegram and social betting; (6) cross-chain tokens; and (7) marketplace tokens such as Airbnb, buying and selling, and rental platforms.

Some of the reasons given by respondents for their investment choices provides an insight into how perspectives may differ regarding the type of ICOs. Interestingly, the majority believe that currency coins are the most liquid option to invest and return a short-term profit while tokens are associated with startups and projects that either have long-term return or are doomed to fail. While in minority, the other types of use-case tokens also have their own advocates such as the alternative energy platforms to reduce carbon emissions and social platforms due to expected high value returns.

Notably, currency coins such as Bitcoin, Binance, and NEX are the most chosen alternatives as compared to the other types of ICO tokens. Consequently, two alternatives have been modelled, namely the choice of investing in coins and the choice of investing in other types of ICO tokens. The reason for this grouping is that some blockchain advocates focus more on bitcoin and other coins as an alternative to fiat money, while others focus more on the use-cases of blockchain other than the concept of cryptocurrency.

The estimates presented in Exhibit 4 demonstrate the differential effect of these parameters on the utility of coins compared to the other ICO tokens. Accordingly, the Australian model provides evidence of interest in coin investment by those who have a high income, have studied IT, work as freelancers or in the wholesale sector, work in casual employment or are unemployed. However, people with the age of 18 to 30 prefer other use-cases of cryptocurrency. The Chinese survey reveals that male investors, freelancers, and those who have an investment in equity and/or fixed income prefer investment in coins.

Exhibit 4: Statistics of variables – ICO categories

| Alternatives | | Australian survey | | | Chinese Survey | | |
|---|---|---|---|---|---|---|---|
| | Parameters | Statistics | | Parameters | Statistics | |
| | | Average | t - test | | Average | t - test |
| **Alternative 1: Coin** | $\beta_{constant}$ | 12.40 | 60.80 | $\beta'_{constant}$ | -2.52 | -7.64 |
| | $\beta_{Age18\_30}$ | -11.60 | -25.80 | $\beta'_{Male}$ | 0.73 | 2.40 |
| | $\beta_{IncomeCat4}$ | 2.64 | 9.79 | $\beta'_{InvestEquiyFixed}$ | 0.98 | 3.63 |
| | $\beta_{Freelancer}$ | 2.15 | 29.90 | $\beta'_{Freelancer}$ | 1.49 | 3.76 |
| | $\beta_{JobWholesale}$ | 0.94 | 2.89 | | | |
| | $\beta_{MajorIT}$ | 2.57 | 15.80 | | | |
| | $B_{CasulaUnemployed}$ | 0.71 | 4.18 | | | |
| **Number of parameters** | | 7 | | 4 | | |
| **Number of observations** | | 78 | | 278 | | |
| **Null LL** | | -54.065 | | -297.087 | | |
| **LL** | | -18.326 | | -136.449 | | |
| **Adjusted $\rho^2$** | | 0.661 | | 0.541 | | |
| **Akaike Information Criterion** | | 50.652 | | 280.898 | | |
| **Bayesian Information Criterion** | | 67.149 | | 295.408 | | |

## 5. Further discussions

Generally, the main reasons that ICOs have been widely recognized as a disruptive and revolutionary technological innovation are its fundraising attributes of easy accessibility, less regulation, and lower costs (Chanson, et al., 2018). Traditionally, companies of different sizes and developing stages need to target different fundraising strategies with a variety of intermediaries and complex regulations. Revolutionarily, an ICO offers a peer-to-peer fundraising process through globally distributed share offerings with nearly no regulations or intermediaries (Chanson, et al., 2018). These unique features enable it to significantly simplify the traditional fundraising process and allow it to contain any or all of the fundraising attributes included in IPOs, venture capital, crowdfunding, and so on (Tapscott and Tapscott, 2017).

However, innovation is always accompanied by problems. The association between cryptocurrency and extensive criminal activities have been widely identified in the literature (Aldridge and Décary-Hétu, 2016; Foley, Karlsen, & Putniņš, 2019; Forgang, 2019). Additionally, while the unique features of ICOs as a fundraising process for companies to improve the efficiency of the fundraising process, it has also made way for massive scams. One of the key challenges with ICOs and the issued tokens is that it is an emerging market where there are significant variances of quality between individual offerings, and there is an absence of a best practice framework to identify good quality from bad and an avoidance of any form of fiduciary responsibility and clarity on consumer protection by issuers (Sehra, et al., 2017). The Satis Crypto-Asset Market Coverage Initiation indicated that over 80% of ICO fundraising activities were scams, and only 4% of the total number of ICOs had succeeded in raising funds (Dowlat and Hodapp, 2018).

These problems have pressured authorities across the globe to take action in order to protect investors and provide clarity, and this eventually led to two significant drops in the price of cryptocurrencies; one at the end of 2017, and another at the beginning of 2018 (Moran, 2018). The statistical result on CoinMarketCap [1] indicates that the cryptocurrency market capitalisation has fallen by approximately 85% compared to its peak value. These events have had a devastating impact upon the cryptocurrency market and have certainly made the fundraising and financing process of blockchain-related start-ups more difficult. Although the total amount of funds raised is increasing, the overall ICO success rate and fundraising for each project have both been reducing according to ICO Market Research reports on different quarters in 2018[1].

---

[1] CoinMarketCap. (2019). Global Chartes Total Market Capitalization. Retrieved Date from https://coinmarketcap.com/charts/.

*5.1. Risk factors*

While cryptocurrency displays a great deal of potential, the way it has been treated as an investment has caused the market to lose confidence in its longevity in the future. The continuing decline of cryptocurrency prices at the end of 2017 and beginning of 2018 made investors rethink the fundamental nature of the market and its underlying issues, prompting the market to transform from a speculative bubble to a mature market.

To put things into perspective, since its peak at the end of 2017, the cryptocurrency market capitalisation has fallen by approximately 85%, leading many to believe the market to be a purely speculative bubble and may be doomed to disappear in the future (Zetzsche, et al., 2017). Accordingly, we asked the survey participants to rank risk factors when it comes to making decisions about investment in cryptocurrency. As presented in Exhibit 5, there is a slight difference between the Chinese and Australian survey. For instance, Australian are far more concerned about face ICO and blockchain-based start-ups compared to Chinese investors.

Interestingly, regulatory challenges are perceived with less risk among Chinese investors relative to Australians. The regulatory challenges are largely relevant to ICOs given that their regulatory classification can have profound impacts upon profit margins of investors, thereby affecting how well an ICO will perform. The lack of ICO regulation has caused the market to lose confidence in the legitimacy of ICOs, and the survey results confirm this statement.

---

[1] ICO Market Research Quarter 1, Quearter 2, and Quearter 3 2018.

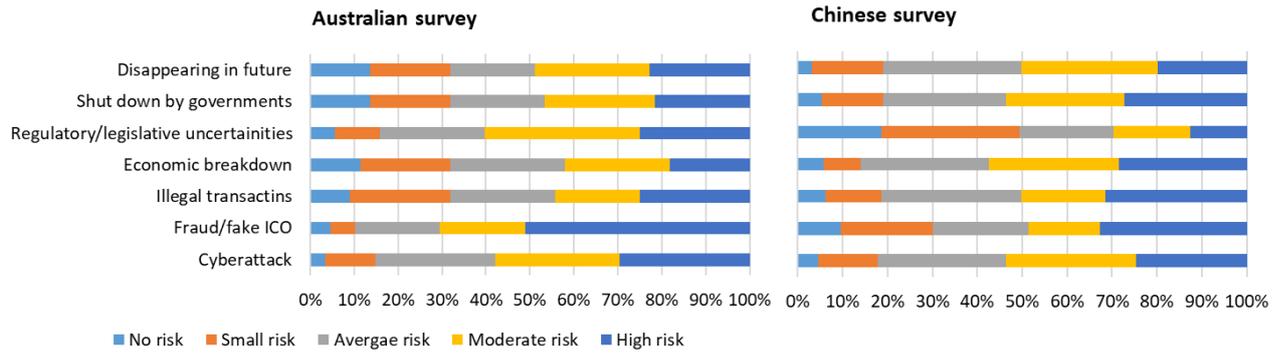

Exhibit 5: statistics of risk factors

## *5.2. ICO attributes*

Survey participants were also asked about the importance of ICO attributes in investment decisions. As shown in Exhibit 6, it turns out that there are a few differences between Australian and Chinese investors. For example, Australians highly value viability factors such as having a long-term strategic plan, white paper, use-case, user-friendly interface, and regional scale when making decisions to invest in an ICO. On the other hand, Chinese individuals highly value the type of platform but care less about the regional scalability and white paper. This may imply that in our sample, Australians are relatively long-term investors as compared to the Chinese.

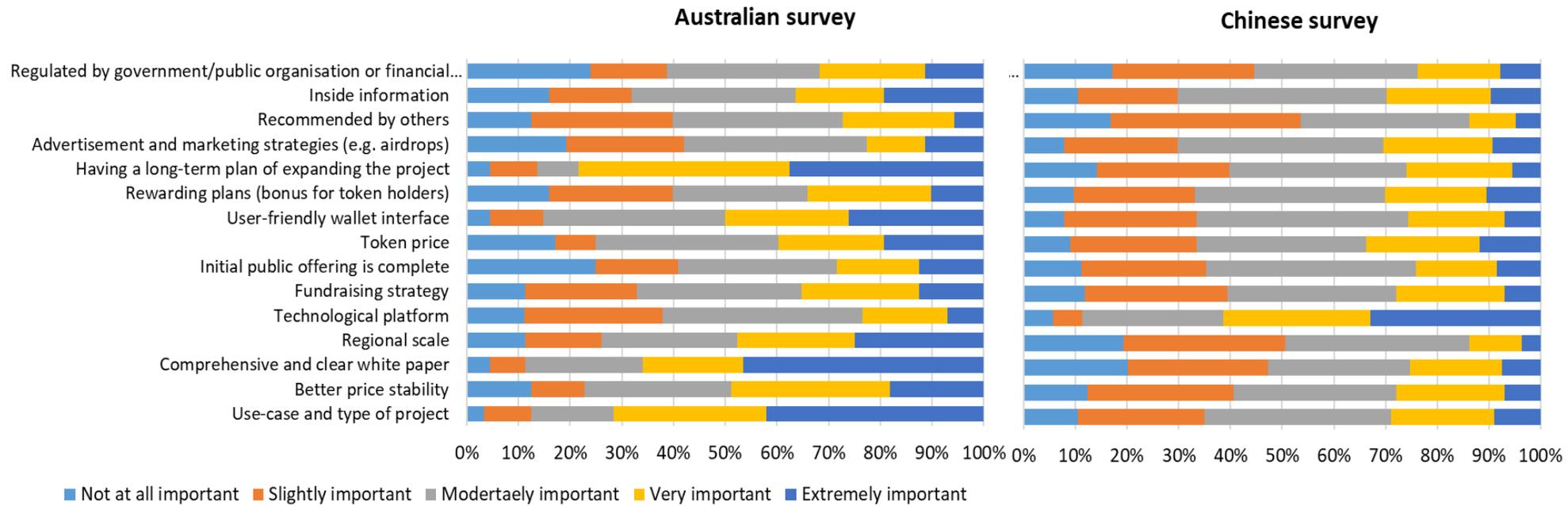

Exhibit 6: ICO attributes

*5.3. Investment strategy*

Respondents were also asked to reveal their investment strategies. As shown in Exhibit 7, the patterns are relatively similar among Australian and Chinese investors, with no changes and increasing across multiple cryptocurrencies as the major strategies.

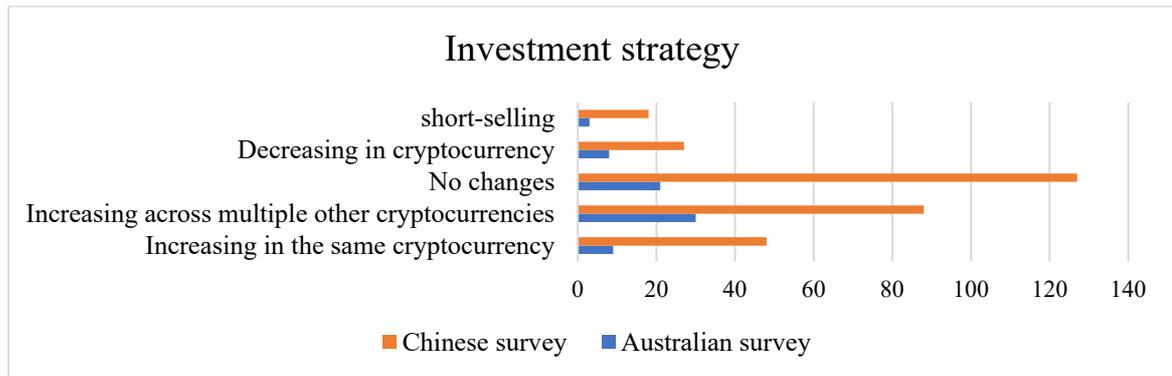

Exhibit 7: Investment strategy

*5.4. Cryptocurrency marketplace*

Exhibit 8 presents the common marketplaces for trading cryptocurrency, as expressed by the Australian and Chinese respondents. While the Chinese use relatively all means of trading, Australian respondents primarily use the online exchange platforms.

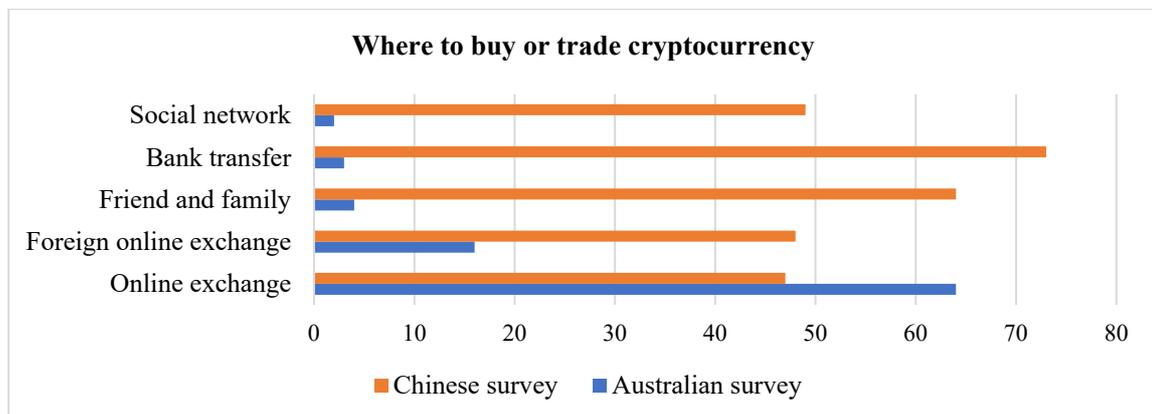

Exhibit 8: Common marketplaces to trade cryptocurrency

*5.5. Market followers and deterrents*

As mentioned above, the survey was distributed among the followers of cryptocurrency social media channels (e.g. Telegram, facebook, slacks, and etc.) and attendees of cryptocurrency community meetup groups. Notably, while all surbey respondents are closely following the cryptocurrency news, some of them have not invested yet in the market. Hence, we asked this group about the main detterent reasons that have made them not to invest.

Interestingly, there are differences between the Chinese and Australian responses due to different regulations in the two countries, as shown in Exhibit 9. The main reasons mentioned by Chinese folks are lack of enough knowledge, volatility compared to fiat money, and failure stories. Meanwhile, the main factors for Australians are insufficient regulations, lack of enough information, concerns about trading, and uncertainty over the technology's future.

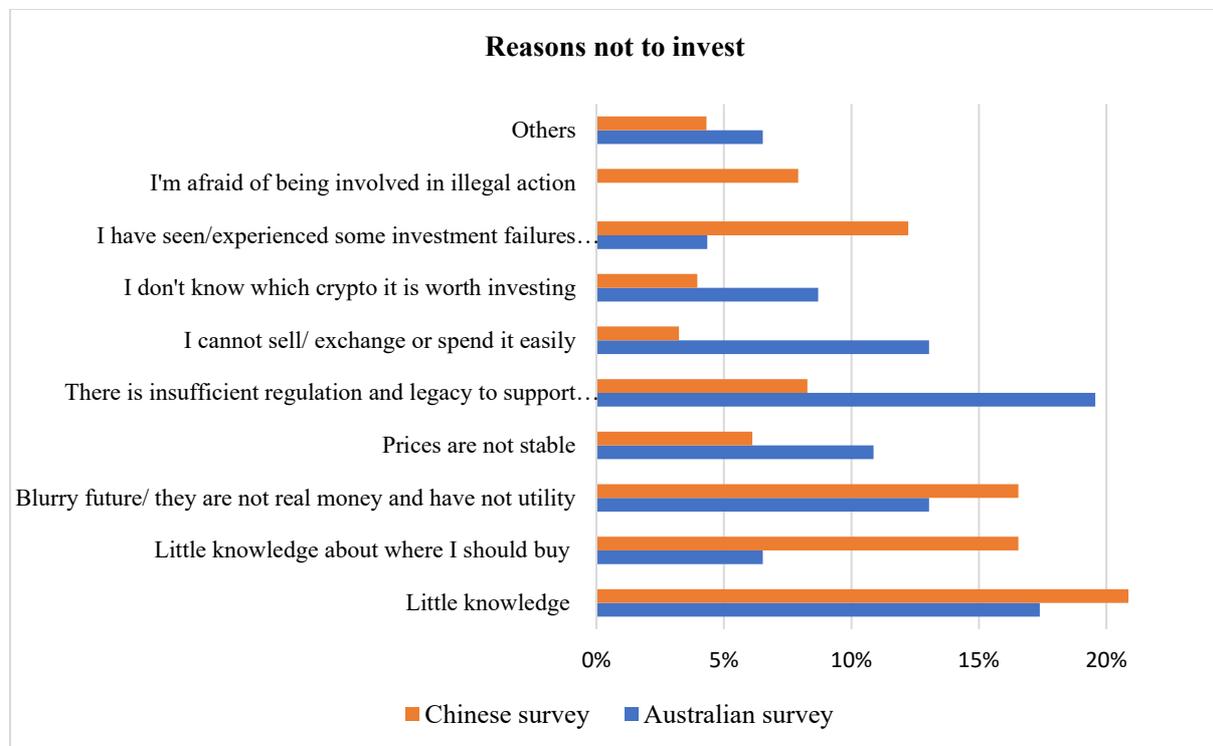

Exhibit 9: Deterrents of cryptocurrency

## 6. Conclusion

The results of this research have produced findings relating to gender, income, age, occupation, investment experience, and further inferences. What is interesting about these initial findings is just how well they align with theories of risk-aversion in behavioral economics and finance. Evidence suggests that less Australian females have invested in cryptocurrency up to now and are more likely to never invest or invest in the future. Research in behavioral finance and economics suggests that where men tend to be more overconfident, women are much more cautious in situations of significant uncertainty. This could be one possible explanation of our findings given that the cryptocurrency market has experienced significant fluctuation with a very blurry future given recent trends in pricing and regulation.

As would be expected, income is also a significant determinant of cryptocurrency investment amongst Australians, with those who have already invested in the market having incomes between 1000 and 2500 AUD per week. This is in line with behavioral research which says we become more risk tolerant as our income increases, and it is also in line with the nature of crypto which allows people who have normally never invested beyond their superannuation to invest in something that gives them a chance to get "rich."

Amongst Chinese individuals, the probability of investing in cryptocurrency increases for individuals belonging to the age group of 18-30 years old. This relationship is in line with biological scientists finding that the risk-seeking part of our brains physically changes as we age. It could also be linked with the fact that young people are more exposed and familiar with such disruptive technologies, or advocate the idea of replacing fiat money. The survey reveals that Chinese individuals who are older than 40 intend to invest in the future, which may imply an increase in the expendable income of this age group in China, thereby shifting their values.

It seems that further knowledge about cryptocurrency and economics impacts the decision of investment as Australian respondents who have studied economics, business, or finance are more likely to invest in the future, and having a postgraduate degree has been found to be a positive predictor of cryptocurrency investment for Chinese individuals. Through this analysis, it has also been found that Australians who have invested or intend to invest in the cryptocurrency have jobs in the banking and IT sectors. Those in banking could be expected to be interested in cryptocurrency investment as a reflection of how the sector thrives upon financial innovation. Additionally, given the inherent expertise and resources those in IT may have over the general population, it is reasonable to expect that they may be more likely to exhibit interest in cryptocurrency investment. However, Australian business owners, entrepreneurs, and those who are employed in the education sector are less likely to invest in cryptocurrency in the future. Furthermore, Chinese individuals who are employed in the wholesale sector are unlikely to invest in the cryptocurrency market.

Australian share market investors are less likely to invest in cryptocurrency in the future. This could be due to the market crash in 2017 causing them to sell or cut their losses which could have also led them to question the market's fundamental value, potentially leading them to withdraw and continue investing in more established markets.

This study addressed another research question about the type of token investment, examining the characteristics of investors in coins versus other types of tokens. Looking at the results, Australian investors at the age of 18 to 30 years old are more likely to invest in other startup ICOs compared to coins. However, high-income earners (income higher than 2,500 AUD per week), wholesalers, freelancers, casual workers, the unemployed, and those who have studied IT are more likely to invest in coins. For Chinese investors, these parameters are slightly different with being male, having previously invested in shares and fixed income, and freelancing as the significant, positive predictors for in investing in coins.

Despite the rapid growth of ICOs and skyrocketing cryptocurrency valuations, the market is yet to reach a maturity level where it is feasible to be used by corporate funds alongside individuals. According to Gartner, ICOs are on the peak of a hype cycle while cryptocurrencies are passing a trough of disillusionment. A significant issue within the market is irrational investor behavior. One example is herd mentality which occurs when individuals disregard private information in favour of a general consensus. Many other biases are evident, such as overconfidence and the disposition effect, indicating the market to be filled with irrational decision-making, ultimately leading to an inefficient market with a poor pricing mechanism.

Behavioral issues such as these could begin to be corrected by introducing professional investors to the market due to their experience in equity research. However, governments around the world have had different reactions to the innovative technology, with countries such as China outright banning cryptocurrency trade through to countries like America which have begun applying similar regulations to cryptocurrency as they do to equities and other securities. Due to this global division and regulatory uncertainty, it is unlikely that institutional investors will enter the market in any meaningful way in the near future.

However, ICOs and cryptocurrency may become mainstream in the future. Grounded on the concept of economy fragmentation which bitcoin advocated, it is important to study what other socio-demographic characteristics and cryptocurrency attributes there are which may affect individuals (rather than corporate funds) to participate and invest in the market. Hence, this study can be further extended by incorporating latent variables such as risk, social beliefs, and technology enthusiasm.

## 7. References

Adhami, S., Giudici, G., & Martinazzi, S. (2018). Why do businesses go crypto? An empirical analysis of initial coin offerings. *Journal of Economics and Business, 100*, pp. 64-75. doi:10.1016/j.jeconbus.2018.04.001

Akerlof, G. A. (1970). The Market for "Lemons": Quality Uncertainty and the Market Mechanism. *The Quarterly Journal of Economics, 84*(3), pp. 488-500. Retrieved from http://www.utdallas.edu/~nina.baranchuk/Fin7310/papers/Akerlof1970.pdf

Alam, S. (2017). Testing the weak form of efficient market in cryptocurrency. *Journal of Engineering and Applied Sciences, 12*(9), pp. 2285-2288. Retrieved from http://docsdrive.com/pdfs/medwelljournals/jeasci/2017/2285-2288.pdf

Aldridge, J., & Décary-Hétu, D. (2016). Hidden wholesale: The drug diffusing capacity of online drug cryptomarkets. *International Journal of Drug Policy, 35*, pp. 7-15. doi:10.1016/j.drugpo.2016.04.020

Allen, F., & Faulhaber, G. R. (1989). Signalling by Underpricing in the IPO Market. *Journal of Financial Economics, 23*, pp. 303-323. doi:10.1016/0304-405X(89)90060-3

Almenberg, J., & Dreber, A. (2015). Gender, stock market participation and financial literacy. *Economics Letters, 137*, pp. 140-142. doi:10.1016/j.econlet.2015.10.009

Barber, B. M., & Odean, T. (2001). Boys Will Be Boys: Gender, Overconfidence, and Common Stock Investmen. *The Quarterly Journal of Economics*doi:10.1162/003355301556400

Bariviera, A. F., ́e, W. H., Basgall, M. ı. J. e., & Naiouf, M. (2017). Some stylized facts of the Bitcoin market. *Physica A: Statistical Mechanics and its Applications, 484*, pp. 82-90. doi:10.1016/j.physa.2017.04.159

Barsan, I. M., & LL.M. (2017). Legal Challenges of Initial Coin Offerings (ICO). *Available at SSRN*Retrieved from https://ssrn.com/abstract=3064397

Bartos, J. (2015). Does Bitcoin follow the hypothesis of efficient market? *International Journal of Economic Sciences, IV*(2), pp. 10-23. doi:10.20472/es.2015.4.2.002

Bian, S., Deng, Z., Li, F., Monroe, W., Shi, P., Sun, Z., Wu, W., Wang, S., Wang, W. Y., Yuan, A., Zhang, T., & Li, J. (2018). IcoRating: A Deep-Learning System for Scam ICO Identification. *arXiv preprint aXiv:1803.03670*Retrieved from https://arxiv.org/abs/1803.03670

Bierlaire, M. (2018). *PandasBiogeme: a short introduction*. Transport and Mobility Laboratory, ENAC.

Boreiko, D., & Vidusso, G. (2019). New blockchain intermediaries: do ICO rating websites do their job well? *The Journal of Alternative Investments, 21*(4), pp. 67-79. doi:10.3905/jai.2019.21.4.067

Bourveau, T., George, E. T. D., Ellahie, A., & Macciocchi, D. (2018). Initial Coin Offerings: Early Evidence on the Role of Disclosure in the Unregulated Crypto Market. *Available at SSRN*doi:10.2139/ssrn.3193392

Bucciol, A., & Zarri, L. (2015). The shadow of the past: Financial risk taking and negative life events. *Journal of Economic Psychology, 48*, pp. 1-16. doi:10.1016/j.joep.2015.02.006

Byrnes, J. P., Miller, D. C., & Schafer, W. D. (1999). Gender differences in risk taking: A meta-analysis. *Psychological Bulletin, 125*(3), pp. 367-383. doi:10.1037/0033-2909.125.3.367

Caporale, G. M., Gil-Alana, L., & Plastun, A. (2018). Persistence in the cryptocurrency market. *Research in International Business and Finance, 46*, pp. 141-148. doi:10.1016/j.ribaf.2018.01.002

Carducci, B. J., & Wong, A. S. (1998). Type a and Risk Taking in Everyday Money Matters. [journal article]. *Journal of Business and Psychology, 12*(3), pp. 355-359. doi:10.1023/a:1025031614989 Retrieved from https://doi.org/10.1023/A:1025031614989

Chanson, M., Gjoen, J., Risius, M., & Wortmann, F. (2018) *Initial coin offerings: the role of social media for organizational legitimacy and underpricing*. Paper presented at the International Conference on Information Systems (ICIS), San Francisco, CA, USA. https://www.alexandria.unisg.ch/255399/

Charness, G., & Gneezy, U. (2012). Strong Evidence for Gender Differences in Risk Taking. *Journal of Economic Behavior & Organization, 83*(1), pp. 50-58. doi:10.1016/j.jebo.2011.06.007

Chod, J., & Lyandres, E. (2018). A Theory of ICOs Diversation, Agency, and Information Asymmetry.

Connelly, B. L., Certo, S. T., Ireland, R. D., & Reutzel, C. R. (2010). Signaling Theory: A Review and Assessment. *Journal of Management, 37*(1), pp. 39-67. doi:10.1177/0149206310388419

Courtney, C., Dutta, S., & Li, Y. (2017). Resolving Information Asymmetry: Signaling, Endorsement, and Crowdfunding Success. *Entrepreneurship Theory and Practice, 41*(2), pp. 265-290. doi:10.1111/etap.12267

Deo, M., & Sundar, V. (2015). Gender Difference: Investment Behavior and Risk Taking. *Journal of Indian Management*

Dowlat, S., & Hodapp, M. (2018). *CRYPTOASSET MARKET COVERAGE INITIATION: NETWORK CREATION*. Satis Group: https://research.bloomberg.com/pub/res/d28giW28tf6G7T_Wr77aU0gDgFQ

Felix, T. (2018). Underpricing in the Cryptocurrency World: Evidence from Initial Coin Offerings. *Available at SSRN* doi:10.2139/ssrn.3202320

Feng, C., Li, N., Lu, B., Wong, M. H. F., & Zhang, M. (2018). Initial Coin Offerings, Blockchain Technology, and White Paper Disclosures. *SSRN*

Finke, M. S., & Huston, S. J. (2003). The Brighter Side of Financial Risk: Financial Risk Tolerance and Wealth. [journal article]. *Journal of Family and Economic Issues, 24*(3), pp. 233-256. doi:10.1023/a:1025443204681 Retrieved from https://doi.org/10.1023/A:1025443204681

Firoozi, F., Jalilvand, A., & Lien, D. (2017). Information Asymmetry and Adverse Wealth Effects of Crowdfunding. *The Journal of Entrepreneurial Finance, 18*(1)Retrieved from http://hdl.handle.net/10419/197551

Fisch, C. (2019). Initial coin offerings (ICOs) to finance new ventures. *Journal of Business Venturing, 34*(1), pp. 1-22. doi:10.1016/j.jbusvent.2018.09.007

Fisher, P. J., & Yao, R. (2017). Gender differences in financial risk tolerance. *Journal of Economic Psychology, 61*, pp. 191-202. doi:10.1016/j.joep.2017.03.006

Foley and Lardner LLP. (2018). *Cryptocurrency Survey*. https://www.foley.com/files/uploads/Foley-Cryptocurrency-Survey.pdf

Foley, S., Karlsen, J. R., & Putniņš, T. J. (2019). Sex, drugs, and bitcoin: How much illegal activity is financed through cryptocurrencies? *The Review of Financial Studies, 32*(5), pp. 1798-1853.

Forgang, G. (2019). Money Laundering Through Cryptocurrencies. *Economic Crime Forensics Capstones* Retrieved from https://digitalcommons.lasalle.edu/ecf_capstones/40/

Grable, J. E. (2000). Financial Risk Tolerance and Additional Factors That Affect Risk Taking in Everyday Money Matters. [journal article]. *Journal of Business and Psychology, 14*(4), pp. 625-630. doi:10.1023/a:1022994314982 Retrieved from https://doi.org/10.1023/A:1022994314982

Howell, S. T., Niessner, M., & Yermack, D. (2018). Initial coin offerings: Financing growth with cryptocurrency token sales. *The National Bureau of Economic Research* Retrieved from https://www.nber.org/papers/w24774

Jacobsen, B., Lee, J. B., Marquering, W., & Zhang, C. Y. (2014). Gender differences in optimism and asset allocation. *Journal of Economic Behavior & Organization, 107*, pp. 630-651. doi:10.1016/j.jebo.2014.03.007

Kromidha, E., & Robson, P. (2016). Social identity and signalling success factors in online crowdfunding. *Entrepreneurship & Regional Development, 28*(9-10), pp. 605-629. doi:10.1080/08985626.2016.1198425

Kurihara, Y., & Fukushima, A. (2017). The Market Efficiency of Bitcoin: A Weekly Anomaly Perspective. *Journal of Applied Finance & Banking, 7*(3), pp. 57-64. Retrieved from http://www.scienpress.com/Upload/JAFB%2FVol%207_3_4.pdf

Latif, S. R., Mohd, M. A., Mohd Amin, M. N., & Mohamad, A. I. (2017). Testing the Weak Form of Efficient Market in Cryptocurrency. *Journal of Engineering and Applied Sciences, 12*(9), pp. 2285-2288.

Lemaster, P., & Strough, J. (2014). Beyond Mars and Venus: Understanding gender differences in financial risk tolerance. *Journal of Economic Psychology, 42*, pp. 148-160. doi:10.1016/j.joep.2013.11.001

Mahomed, N. (2017). *Understanding consumer adoption of cryptocurrencies* (Master of Business Administration. University of Pretoria.

Momtaz, P. P. (2018). Initial Coin Offerings, Asymmetric Information, and Loyal CEOs. *Available at SSRN*doi:10.2139/ssrn.3167061

Moran, J. D. (2018). The Impact of Regulatory Measures Imposed on Initial Coin Offering. *Catholic University Journal of Law and Technology, 26*(2)Retrieved from http://citeseerx.ist.psu.edu/viewdoc/summary?doi=10.1.1.221.9986

Morin, R. A., & Suarez, A. F. (1983). Risk Aversion Revisited. *The Journal of Finance, 4*, pp. 1201-1216. Retrieved from https://www.jstor.org/stable/2328020

Nakamoto, S. (2008). Bitcoin: A peer-to-peer electronic cash system.

Ndirangu, A. W., Ouma, B. O., & Munyaka, F. G. (2014). Factors Influencing Individual Investor Behaviour during Initial Public Offers (IPOs) in Kenya. *II*(8)

Neelakantan, U. (2010). Estimation and Impact of Gender Differences in Risk Tolerance. *Economic Inquiry, 48*(1), pp. 228-233. doi:10.1111/j.1465-7295.2009.00251.x

Riley, W. B., & Chow, K. V. (1992). Asset Allocation and Individual Risk Aversion. *Financial Analysts Journal, 48*(6), pp. 32-37. doi:10.2469/faj.v48.n6.32 Retrieved from https://doi.org/10.2469/faj.v48.n6.32

Roszkowski, M. J., & Grable, J. E. (2010). Gender differences in personal income and financial risk tolerance: How much of a connection? *The Career Development Quarterly, 58*(3), pp. 270-275.

Sehra, A., Smith, P., & Gomes, P. (2017). Economics of Initial Coin Offerings.

Spence, M. (1978). Job market signaling *Uncertainty in economics* (pp. 281-306): Elsevier.

Sung, J., & Hanna, S. D. (1996). Factors related to household risk tolerance: An ordered probit analysis. *Sung, J. & Hanna, S.(1996). Factors related to household risk tolerance: An ordered probit analysis. Consumer Interests Annual, 42*, pp. 227-228.

Tapscott, A., & Tapscott, D. (2017). How Blockchain Is Changing Finance. *Harvard Business Review*Retrieved from https://www.bedicon.org/wp-content/uploads/2018/01/finance_topic2_source2.pdf

Train, K. E. (2009). *Discrete choice methods with simulation*: Cambridge university press.

Urquhart, A. (2016). The inefficiency of Bitcoin. *Economics Letters, 148*, pp. 80-82. doi:10.1016/j.econlet.2016.09.019

Wang, H., & Hanna, S. (1997). Does Risk tolerance Decrease with Age? *Financial Counseling and Planning, 8*(2)Retrieved from https://papers.ssrn.com/sol3/papers.cfm?abstract_id=95489

Zetzsche, D. A., Buckley, R. P., Arner, D. W., & Föhr, L. (2017). The ICO Gold Rush: It's a scam, it's a bubble, it's a super challenge for regulators. *University of Luxembourg Law Working Paper*(11), pp. 17-83. doi:10.2139/ssrn.3072298

Zhou, W.-X., Kim, Y. B., Kim, J. G., Kim, W., Im, J. H., Kim, T. H., Kang, S. J., & Kim, C. H. (2016). Predicting Fluctuations in Cryptocurrency Transactions Based on User Comments and Replies. *PloS one, 11*(8)doi:10.1371/journal.pone.0161197

# Appendix 1: Characteristics of the survey participants

| **Australian survey** | |
|---|---|
| **Characteristics** | **Distribution** |
| Gender | Male (65%); female (34%); other (1%) |
| Age | 18-30 years (44%); 31-40 years (18%); 41-50 years (19%); 51-60 years (14%); 60 years and older (5%) |
| Employment status | Full-time employee (36%); student (26%); business owner (15%); freelancer (12%); academic staff/researcher (7%); casual worker (2%); unemployed (2%) |
| Job sector | IT (34%); engineering and/or health specialists/professionals (16%); entrepreneur and business owner (14%); banking (9%); Education (8%); retail sales, marketing and advertising professional (4%); wholesale trade, producer, manufacturer, distribution and logistics (4%); journalist, reporter, author (1%); other (10%) |
| Education level | Year-12 certificate (15%); vocational education (15%); undergraduate (34%); postgraduate (36%) |
| Major of study | Economics (35%); finance (33%); IT and computer science/software engineering (8%); science (5%); business (5%); other engineering (7%); law (2%); psychology/human sciences (3%); other (2%) |
| Weekly income level | Above 2500 AUD (16%); 1000-2500 AUD (35%); 650-1000 AUD (9%); 0-650 AUD (25%); Negative income (15%) |
| Investment experience | Equities e.g. public-listed shares (22%); fixed income e.g. government and corporate bonds (12%); cash e.g. saving accounts and term deposits (15%); property e.g. house (35%); none (15%) |
| **Chinese survey** | |
| **Characteristics** | **Distribution** |
| Gender | Male (38%); female (59%); other (3%) |
| Age | 18-30 years (62%); 31-40 years (28%); 41-50 years (9%); 51-60 years (1.0%); 60 years and older (0%); |
| Employment | Full-time employee (36%); student (17%); business owner (8%); freelancer (14%); academic staff/researcher (14%); casual worker (6%); unemployed (5%) |
| Job sector | IT (16%); engineering, health specialists, professionals (2%); entrepreneur and business owner (7%); banking (3%); education (16%); retail sales, marketing and advertising professional (20%); wholesale trade, producer, manufacturer, distribution and logistics (14%); journalist/reporter/author (5%); other (17%). |
| Education | Year-12 certificate (13%); vocational education (23%); undergraduate (52%); postgraduate (12%) |
| Major of study | Economics (9%); finance (15%); IT and computer science/software engineering (13%); science (8%); business (11%); other engineering (15%); law (10%); psychology/human sciences (6%); other (8%); no study major (5%). |
| Weekly income level | Above ¥9300 (3%); ¥3720-¥9300 (13%); ¥2418-¥3720 (32%); ¥0-¥2418 (37%); negative income (15%) |
| Investment options | Equities e.g. public-listed shares (25%); fixed income e.g. government and corporate bonds (31%); cash e.g. saving accounts and term deposits (11%); property e.g. house (14%); none (19%) |